\documentclass[aps, pra, reprint, floatfix, superscriptaddress, showkeys, nofootinbib, fleqn]{revtex4-2}
\pdfoutput=1

\usepackage{mathtools, amsfonts, amssymb, amsthm, bm, bbm}
\usepackage[sort&compress]{natbib}
\usepackage[colorlinks=true, linkcolor=blue, citecolor=magenta, urlcolor=blue]{hyperref}
\usepackage{lmodern}
\usepackage[T1]{fontenc}
\usepackage[utf8]{inputenc}
\usepackage[american]{babel}
\usepackage[activate={true,nocompatibility},final,tracking=true,kerning=true,spacing=true,factor=1100]{microtype}
\usepackage{orcidlink}
\usepackage{multirow, booktabs}
\usepackage{nccmath}
\usepackage[margin=1.8cm]{geometry}
\usepackage{graphics}
\usepackage{titlesec}

\theoremstyle{remark}
\newtheorem{rmk}{Remark}
\newtheorem{ex}{Example}
\newtheorem*{rmk*}{Remark}
\newtheorem*{ex*}{Example}

\newcommand{\HH}{\mathcal{H}}
\newcommand{\RR}{\mathcal{R}}
\newcommand{\GG}{\mathcal{G}}
\newcommand{\VV}{\mathcal{V}}
\newcommand{\UU}{\mathrm{U}}
\newcommand{\uu}{\mathfrak{u}}
\renewcommand{\AA}{\mathcal{A}}
\newcommand{\tr}{\operatorname{tr}}
\renewcommand{\d}{\mathrm{d}}
\newcommand{\dt}{\mathrm{d}t}
\newcommand{\ds}{\mathrm{d}s}

\newcommand{\bra}[1]{\langle #1|}
\newcommand{\ket}[1]{|#1\rangle}
\newcommand{\braket}[2]{\langle #1 | #2 \rangle}
\newcommand{\ketbra}[2]{|#1 \rangle\langle #2|}
\newcommand{\llangle}{\langle\!\langle}
\newcommand{\rrangle}{\rangle\!\rangle}

\newcommand{\gG}{g_{\mathcal{G}}}
\newcommand{\gV}{g_{\mathcal{V}}}

\newcommand{\hol}{\Gamma}
\newcommand{\dyn}{\mathrm{D}}
\newcommand{\qsl}{\operatorname{\tau}}
\newcommand{\diag}{\operatorname{diag}}
\newcommand{\length}{\mathcal{L}}
\newcommand{\energy}{\mathcal{E}}

\begin{document}
\title{Estimate of the time required to perform a nonadiabatic \\ holonomic quantum computation}
\author{Ole S{\"o}nnerborn\,\orcidlink{0000-0002-1726-4892}\,}
\email{ole.sonnerborn@kau.se}
\affiliation{Department of Mathematics and Computer Science, Karlstad University, 651 88 Karlstad, Sweden}
\affiliation{Department of Physics, Stockholm University, 106 91 Stockholm, Sweden}
\date{\today}

\begin{abstract}
Nonadiabatic holonomic quantum computation has been proposed as a method to implement quantum logic gates with robustness comparable to that of adiabatic holonomic gates but with shorter execution times. In this paper, we establish an isoholonomic inequality for quantum gates, which provides a lower bound on the lengths of cyclic transformations of the computational space that generate a specific gate. Then, as a corollary, we derive a nonadiabatic execution time estimate for holonomic gates. In addition, we demonstrate that under certain dimensional conditions, the isoholonomic inequality is tight in the sense that every gate on the computational space can be implemented holonomically and unitarily in a time-optimal way. We illustrate the results by showing that the procedures for implementing a universal set of holonomic gates proposed in a pioneering paper on nonadiabatic holonomic quantum computation saturate the isoholonomic inequality and are thus time optimal.
\end{abstract}


\maketitle

\titleformat{\section}[hang]{\bfseries\large}{\arabic{section}}{0.9em}{}
\titlespacing{\section}{0em}{1.2em}{0.7em}
\titleformat{\subsection}[block]{\bfseries\normalsize}{\arabic{section}.\arabic{subsection}}{0.7em}{}
\titlespacing{\subsection}{0em}{1em}{0.5em}
\titleformat{\subsubsection}[block]{\itshape\normalsize}{\arabic{section}.\arabic{subsection}.\arabic{subsubsection}}{0.4em}{}
\titlespacing{\subsubsection}{0em}{0.6em}{0.3em}

\section{Introduction}\label{sec: Introduction}
Adiabatic holonomic computation has been launched as a procedure to implement quantum gates resilient to certain types of errors \cite{WiZe1984, PaZaRa1999, ZaRa1999, PaZa2001}. However, the slow parametric control associated with adiabatic evolution makes adiabatic computations sensitive to external perturbations. To address this issue, an alternative method for realizing quantum gates using nonadiabatic holonomies has been proposed \cite{An1988, SjToAnHeJoSi2012, XuZhToSjKw2012, AlSj2022, ZhKyFiKwSjTo2023}.\footnote{The importance of developing nonadiabatic computational schemes has also been emphasized and demonstrated in the related field of geometric quantum computation; see Refs.\ \cite{Wang2001, Zhu2002}.} A nonadiabatic holonomic computation exploits the system’s internal dynamics, which significantly shortens the execution time of the computation compared to the adiabatic case. However, fundamental properties of quantum mechanical systems preclude arbitrarily short execution times for holonomic quantum gates. In this paper, we derive an estimate of the time required to execute a holonomic quantum gate unitarily. This estimate builds upon and generalizes a corresponding estimate of the time it takes to generate an Aharonov-Anandan geometric phase, as reported in Ref.\ \cite{HoSo2023c}.

The main ingredient in the derivation of the execution time estimate is the isoholonomic inequality for quantum gates. The isoholonomic inequality establishes a minimum length for cyclic transformations of the computational space that holonomically generate a specific gate. This inequality, together with the results in Ref.\ \cite{TaNaHa2005}, solves the isoholonomic problem for quantum gates formulated by Montgomery \cite{Mo1990}. 

Holonomic gates are the building blocks of circuits in holonomic quantum computation. Since holonomic gates have a purely geometric origin, implementations of quantum gates through parallel transport operators are predicted to be highly robust against noise \cite{PaZa2001}. Nonadiabatic holonomic gates have been experimentally demonstrated in various physical systems \cite{AbFiJuPeBeWaFi2013, FeXuLo2013, A-CLaHeBa2014, ZuWaHeZhDaWaDu2014}. We show that the scheme in the pioneering paper \cite{SjToAnHeJoSi2012} for the implementation of a universal set of holonomic gates is time optimal.

The paper is organized as follows. Section \ref{sec: Results} presents the main results. Section \ref{sec: Parallel transport and holonomic gates} introduces terminology and describes basic properties of Stiefel-Grassmann bundles. Section \ref{sec: The isoholonomic inequality} contains derivations of the main results. In Sec.\ \ref{sec: Nonadiabatic holonomic computation} we apply the main results to a proposal on how to experimentally implement a universal set of holonomic quantum gates. Finally, in Sec.\ \ref{sec: Tightness of the isoholonomic bound}, we prove that the isoholonomic inequality is tight in a strong sense provided the dimension of the computational space is at most half of the dimension of the Hilbert space. The paper concludes with a summary.

\section{Results}\label{sec: Results}
Throughout, $\RR$ denotes an $n$-dimensional subspace of a finite-dimensional Hilbert space $\HH$. We 
write $P_\RR$ for the orthogonal projection onto $\RR$, and we
use computational terminology and call 
$\RR$ the computational space and 
unitary operators on $\RR$ gates. 
Moreover, we assume all quantities have units such that $\hbar=1$.

When considering a one-parameter family---a curve---of operators, vectors, or subspaces, we assume that the family depends smoothly on the parameter and that the parameter ranges from $0$ to $\tau$. Also, we refer to the parameter as ``time,'' even though it may not represent actual time. 
We say that the curve is closed, cyclic, or a loop when
the initial and final members of the curve are the same.

Inspired by a question from a colleague, Montgomery \cite{Mo1990} formulated the isoholonomic problem for quantum gates as follows: 
Find the shortest cyclic transformation of a subspace whose holonomy is a given gate. 
In this paper, we provide a partial solution to this problem by deriving a lower bound on the length of a cyclic transformation of $\RR$ in terms of its holonomy: Assume $\RR_t$ is a curve of $n$-dimensional subspaces of $\HH$ that starts and ends with $\RR$.
Let $\Gamma$ be the holonomy of $\RR_t$. Then the length of $\RR_t$ is bounded from below by
\begin{equation}\label{eq: the isoholonomic bound}
    L(\Gamma)=\sqrt{\sum_{j=1}^n |\theta_j|\big(2\pi-|\theta_j|\big)},
\end{equation}
where $\theta_1,\theta_2,\dots,\theta_n$ are the principal arguments of the eigenvalues of $\Gamma$.\footnote{The principal argument of a nonzero complex number $z$ is the phase $\theta$ in the interval $(-\pi,\pi]$ for which $z=|z|e^{i\theta}$.} We refer to the length estimate 
\begin{equation}\label{eq: the isoholonomic inequality}
    \length[R_t]\geq L(\Gamma)
\end{equation}
as the isoholonomic inequality and call $L(\Gamma)$ the isoholonomic bound of the gate $\Gamma$.

The isoholonomic inequality is tight when the dimension of $\RR$ is at most half of the dimension of $\HH$:
Let $\Gamma$ be any gate on $\RR$, and $k$ be the number of $1$s in the spectrum of $\Gamma$.
If the codimension of $\RR$ is at least $n-k$, there is a parallel transporting Hamiltonian that drives $\RR$ in a loop with holonomy $\Gamma$ and length $L(\Gamma)$.

From the isoholonomic inequality one can derive an estimate of the time required to drive $\RR$ unitarily in a loop with a given holonomy.
Assume $\RR_t=U_t(\RR)$, where $U_t$ is the time propagator associated with a Hamiltonian $H_t$.
The square of the speed of $\RR_t$ equals 
\begin{equation}\label{eq: skewness measure}
    I(H_t;\RR_t)=-\frac{1}{2}\tr\big([H_t,P_t]^2\big),
\end{equation}
where $P_t$ is the orthogonal projection onto $\RR_t$.
Write $\llangle I(H_t;\RR_t)^{1/2}\rrangle$ for the average speed of $\RR_t$ over the evolution time interval.
The isoholonomic inequality implies that the evolution time is not smaller than
\begin{equation}\label{eq: the runtime bound}
    \qsl[H_t;\Gamma]=\frac{L(\Gamma)}{\llangle\sqrt{I(H_t;\RR_t)}\,\rrangle}.
\end{equation}

The quantity $I(H_t;\RR_t)$ measures the ``skewness'' of  $H_t$ relative to $\RR_t$; see 
Refs.\ \cite{Gi2014, LuSu2020}. 
If the Hamiltonian is time-independent, $H_t=H$, the skewness is a conserved quantity, and the evolution time is lower bounded by
\begin{equation}
    \tau[H;\Gamma]=\frac{L(\Gamma)}{\sqrt{I(H;\RR)}}.
\end{equation} 

We derive the isoholonomic inequality \eqref{eq: the isoholonomic inequality} and the runtime bound \eqref{eq: the runtime bound} in Sec.\ \ref{sec: The isoholonomic inequality}, and we prove the tightness of the isoholonomic inequality in Sec.\ \ref{sec: Tightness of the isoholonomic bound}.

\section{Parallel transport and holonomic gates}\label{sec: Parallel transport and holonomic gates}
Cyclic transformations of $\RR$ correspond to curves in the Grassmann manifold of $n$-dimensional 
subspaces of $\HH$ that start and end at $\RR$.
The Grassmann manifold can be identified with the manifold of orthogonal projections on $\HH$ of rank $n$ by
identifying each $n$-dimensional subspace of $\HH$ with the orthogonal projection onto that subspace.\footnote{This identification is used to induce a topology and a geometry on the Grassmann manifold from the space of Hermitian operators on $\HH$.
} A cyclic transformation of $\RR$ is then represented by a curve of orthogonal projections that starts and ends at $P_\RR$.
We will use the same notation, $\GG(n;\HH)$, for the space of $n$-dimensional subspaces of $\HH$ and the space of orthogonal projection operators of rank $n$ on $\HH$.

\begin{rmk} 
The elements of $\GG(1;\HH)$ represent the pure states of a quantum system modeled on $\HH$. 
In Ref.\ \cite{HoSo2023c}, we derived time estimates for cyclic transformations of pure states in terms of their Aharonov-Anandan geometric phase. 
Here, we generalize one of these to an estimate of the time required to execute a holonomic gate.
\end{rmk}

An $n$ frame in $\HH$ is an ordered sequence of $n$ orthonormal vectors in $\HH$. 
It will prove convenient to represent an $n$-frame $F$ as a row matrix, 
\begin{equation}
    F=\big(\ket{u_1}\,\ket{u_2}\cdots\ket{u_n}\big).
\end{equation}
We will only consider frames of the unspecified but fixed length $n$,
and will, therefore, only write frame when referring to an $n$ frame.

We can act on $F$ with an operator $A$ defined on its span.
The result is the row matrix $AF$ whose elements are the images of the vectors of $F$ under $A$,
\begin{equation}
    AF=\big( A\ket{u_1}\, A\ket{u_2}\cdots A\ket{u_n}\big).
\end{equation}
The matrix $AF$ is a frame if and only if $A$ is an isometry on the span of $F$. 
We can also act on $F$ from the right by an $n\times k$ numerical matrix $M=(m_{ij})$. 
The result is a row matrix $FM$ whose elements are linear combinations of the vectors of $F$:
\begin{equation}
    FM
    =\bigg(\sum_{i=1}^n m_{i1}\ket{u_i}\,\sum_{i=1}^n m_{i2}\ket{u_i}\cdots\sum_{i=1}^n m_{ik}\ket{u_i}\bigg).
\end{equation}
If $k=1$, $FM$ is a linear combination of the vectors in $F$; if $k=n$ and $M$ is unitary, $FM$ is a frame that spans the same subspace as $F$. 

We can also multiply frames by conjugates of frames. 
Depending on how we multiply them, we get either an operator or a matrix of numbers: If 
$F_1 = (\ket{u_1}\,\ket{u_2}\cdots\ket{u_n})$ and $F_2 = (\ket{v_1}\,\ket{v_2}\cdots\ket{v_n})$,
then
\begin{equation}
    F_1F_2^\dagger = \ketbra{u_1}{v_1}+\ketbra{u_2}{v_2}+\dots+\ketbra{u_n}{v_n}
\end{equation}
and
\begin{equation}
    F_1^\dagger F_2 = 
    \begin{pmatrix}
        \braket{u_1}{v_1} & \braket{u_1}{v_2} & \hdots & \braket{u_1}{v_n} \\
        \braket{u_2}{v_1} & \braket{u_2}{v_2} & \hdots & \braket{u_2}{v_n} \\
        \vdots & \vdots & & \vdots \\
        \braket{u_n}{v_1} & \braket{u_n}{v_2} & \hdots & \braket{u_n}{v_n}        
    \end{pmatrix}.
\end{equation}

The frames in $\HH$ form the Stiefel manifold $\VV(n;\HH)$.
If $F$ is a frame, $FF^\dagger$ is the orthogonal projection operator onto the span of $F$. 
Thus, $FF^\dagger$ belongs to $\GG(n;\HH)$. 
The assignment
\begin{equation}
    \VV(n;\HH)\ni F\to FF^\dagger\in \GG(n;\HH)
\end{equation}
is a principal fiber bundle called the Stiefel-Grassmann bundle. 
This bundle has gauge group the group of unitary matrices $\UU(n)$, which
means that frames $F_1$ and $F_2$ project onto the same projection operator if and only if $F_2 = F_1U$ for an $n\times n$ unitary matrix $U$. 
We recommend the two-volume work \cite{KoNo1996} as a reference for the theory of principal fiber bundles.

\subsection{Parallel transport operators}
Parallel transport operators parallel transport frames. To specify what this means we need to introduce a connection on the Stiefel manifold.

Suppose $\dot F$ is a tangent vector at the frame $F$. 
If we represent $\dot F$ as a row matrix of vectors, $F^\dagger \dot F$ is an element of the Lie algebra $\uu(n)$ of $n\times n$ skew-Hermitian matrices. 
We define $\AA$ as the $\uu(n)$-valued connection on $\VV(n;\HH)$ sending $\dot F$ to $F^\dagger \dot F$,
\begin{equation}
    \AA(\dot F) = F^\dagger \dot F.
\end{equation}
Using standard terminology, we say that $\dot F$ is horizontal if $\AA(\dot F) = 0$. 
We also say that a curve of frames $F_t$ is horizontal if all its velocity vectors are horizontal.

Consider a curve of $n$-dimensional subspaces $\RR_t$ in $\HH$ starting with $\RR$. 
Let $P_t$ be the corresponding curve of projection operators. 
According to a fundamental result from the theory of fiber bundles, there exists a unique one-parameter family of isometries 
$\Pi_t:\RR\to\RR_t$ such that for each frame $F$ for $\RR$, the curve $F_t=\Pi_tF$ is a horizontal lift of $P_t$, that is, a horizontal curve of frames projecting onto $P_t$. 
The isometries $\Pi_t$ are the parallel transport operators associated with $\RR_t$; see Ref.\ \cite{KoNo1996} for details.

We can express the parallel transport operators in terms of the horizontal lift $F_t$ as $\Pi_t=F_t F^\dagger$. 
More generally, if $F_t$ is \emph{any} curve of frames projecting onto $P_t$,
\begin{equation}\label{eq: holonomy in terms of lift}
    \Pi_t=F_t \overset{_\leftarrow}{\mathcal{T}}\exp \bigg( -\int_0^t\AA(\dot F_{s})\,\ds \bigg) F_0^\dagger.
\end{equation}
The symbol $\overset{_\leftarrow}{\mathcal{T}}$ indicates that the exponential is forward time ordered. 

\subsection{Holonomic gates}
If $\RR_t$ describes a cyclic transformation of $\RR$, the final parallel transport operator $\Pi_\tau$ maps $\RR$ isometrically onto itself. This operator is called the holonomy of $\RR_t$. 
We will henceforth write $\hol[\RR_t]$ for the holonomy of $\RR_t$. 
A holonomic gate on $\RR$ is a gate implemented as the holonomy of a cyclic transformation of $\RR$.

The parallel transport operators $\Pi_t$ associated with an evolution $\RR_t$ of $\RR$ move every vector $\ket{\psi}$ in $\RR$ in such a way that, at every $t$, the velocity vector of the curve  $\ket{\psi_t}=\Pi_t\ket{\psi}$ is orthogonal to $\RR_t$. Geometrically, this means that the parallel transport operators cause no time-local rotation within $\RR$. A holonomic gate on $\RR$ is thus a consequence solely of the translational motion of $\RR$ in the Grassmann manifold. This observation underlies the hypothetical claim that holonomic gates should be particularly robust against noise and certain types of implementation errors \cite{PaZa2001, ZhKyFiKwSjTo2023}. 

\subsection{Parallel transporting Hamiltonians}
We say that a Hamiltonian is parallel transporting if the associated time propagator 
parallel translates frames for $\RR$. 
For any Hamiltonian $H_t$, we can define a parallel transporting Hamiltonian $\bar{H}_t$ that drives $\RR$ along the same path and at the same speed as $H_t$: Let $P_t$ be the curve of orthogonal projectors generated from $P_\RR$ by $H_t$ and define $\bar{H}_t$ as
\begin{equation}
    \bar{H}_t = 
    H_tP_t + P_tH_t - 2 P_t H_t P_t.
\end{equation}
Then $[\bar{H}_t, P_t]=[H_t,P_t]$, which shows that $\bar{H}_t$ propagates $\RR$ in the same way as $H_t$, and if $F$ is any frame for $\RR$, and $F_t=\bar{U}_tF$, where $\bar{U}_t$ is the time propagator of $\bar{H}_t$, then $F_t^\dagger \bar{H}_tF_t=0$, which shows that $\bar{H}_t$ is parallel transporting.
For a time-independent Hamiltonian $H$, the corresponding parallel transporting Hamiltonian is 
\begin{equation}\label{eq: parallel for constant H}
    \bar{H}_t = e^{-itH}(HP_\RR + P_\RR H - 2 P_\RR H P_\RR)e^{itH}.
\end{equation}
Although $H$ is time-independent, the parallel transporting Hamiltonian need not be time-independent.

\subsection{Dynamical operators}
The total phase acquired during a cyclic unitary evolution of a pure state can be divided into a geometric part (the holonomy) and a dynamic part \cite{AhAn1987}. It was previously believed that a corresponding division was generally not possible for cyclic unitary evolutions of subspaces. However, Yu and Tong \cite{YoTo2023} recently showed that such a division is always possible. 
Here, we derive the result of Yu and Tong using the framework presented above.

Suppose  $H_t$ drives $\RR$ in a loop $\RR_t$. Let $U_t$ be the time propagator associated with $H_t$.
Choose a frame $F$ for $\RR$ and define a curve of frames as $F_t=U_t F$. According to Eq.\ \eqref{eq: holonomy in terms of lift}, the holonomy of the loop is
\begin{equation}\label{eq: holonomy for Hamiltonian}
    \hol[\RR_t]=U_\tau F\, \overset{_\leftarrow}{\mathcal{T}}\exp\Big(i\int_0^\tau F_t^\dagger H_tF_t\,\dt\Big)F^\dagger.
\end{equation}
We define the dynamical operator of $H_t$ on $\RR$ as
\begin{equation}
    \dyn[H_t]=F\,\overset{_\rightarrow}{\mathcal{T}}\exp\Big(-i\int_0^\tau F_t^\dagger H_tF_t\,\dt \Big)F^\dagger,
\end{equation}
where $\overset{_\rightarrow}{\mathcal{T}}$ indicates that the exponential is backward time-ordered.
By Eq.\ \eqref{eq: holonomy for Hamiltonian}, the restriction of $U_\tau$ to $\RR$ decomposes as
\begin{equation}\label{eq: decomposition of time-evolution operator}
    U_\tau\big|_{\RR}=\hol[\RR_t]\dyn[H_t].
\end{equation}

\section{The isoholonomic inequality}\label{sec: The isoholonomic inequality}
We equip the Grassmann and Stiefel manifolds with the Riemannian metrics
\begin{align}
    \gG(\dot P_1,\dot P_2) &= \frac{1}{2}\tr\big(\dot P_1\dot P_2\big),\\
    \gV(\dot F_1,\dot F_2) &= \frac{1}{2}\tr\big(\dot F_1^\dagger\dot F_2+\dot F_2^\dagger\dot F_1\big).
\end{align}
Furthermore, we define the length of a curve of orthogonal projectors $P_t$ and the length of a curve of frames $F_t$ as
\begin{align}
    \length[P_t]&=\int_0^\tau\sqrt{\gG(\dot P_t,\dot P_t)}\,\dt, \label{eq: length of projector curve} \\ 
    \length[F_t]&=\int_0^\tau\sqrt{\gV(\dot F_t,\dot F_t)}\,\dt.
\end{align}
We also define the kinetic energies of $P_t$ and $F_t$ as
\begin{align}
    \energy[P_t]&=\frac{1}{2}\int_0^\tau\gG(\dot P_t,\dot P_t)\,\dt,\\
    \energy[F_t]&=\frac{1}{2}\int_0^\tau\gV(\dot F_t,\dot F_t)\,\dt.
\end{align}
From the Cauchy-Schwarz inequality we get 
\begin{align}
    2\tau \energy[P_t]&\geq \length[P_t]^2,\\
    2\tau \energy[F_t]&\geq \length[F_t]^2,
\end{align}
with the inequalities being equalities if $P_t$ and $F_t$ have constant speeds.

The Stiefel-Grassmann bundle projection is a Riemannian submersion, which means that the 
tangent map of the projection preserves the inner product between horizontal vectors. 
Consequently, the length of a curve in the Grassmannian and the lengths of all of its horizontal lifts are the same. The same is true for the kinetic energy.

\subsection{The isoholonomic inequality for states}
If $n=1$, the Grassmannian is the projective space of density operators representing pure states of quantum systems modeled on $\HH$, 
and $\gG$ is the Fubini-Study metric.

The holonomy of a closed curve of pure states $\rho_t$ multiplies unit vectors over the common initial and final state by a phase factor. The argument of the holonomy is the  Aharonov-Anandan geometric phase of $\rho_t$ \cite{AhAn1987}. The isoholonomic inequality for states says that the Fubini-Study length of $\rho_t$ is bounded from below as follows:
\begin{equation}\label{eq: the isoholonomic inequality for states}
    \length[\rho_t]\geq L(\theta),\quad L(\theta)=\sqrt{|\theta|(2\pi-|\theta|)},
\end{equation} 
where $\theta$ is the principal argument of the holonomy of $\rho_t$ \cite{Mo1990, HoSo2023c}. 
Below, we extend this inequality to an estimate of the length of a closed curve of subspaces of $\HH$ of arbitrary dimension in terms of the holonomy of the curve. The derivation uses the estimate \eqref{eq: the isoholonomic inequality for states}.
For convenience, we have included a slightly rewritten version of the derivation of the estimate \eqref{eq: the isoholonomic inequality for states} found 
in Ref.\ \cite{HoSo2023c} in Appendix \ref{app: Derivation of the isoholonomic inequality for states}.

\begin{ex}
Consider a qubit with Hamiltonian
\begin{equation}
    H=\epsilon_0\ketbra{0}{0}+\epsilon_1\ketbra{1}{1},\quad \epsilon_0<\epsilon_1.
\end{equation}
Assume the qubit is initially in the pure state $\rho$ and evolves unitarily as $\rho_t=e^{-itH}\rho e^{itH}$. Let $\ket{\psi}=a\ket{0}+b\ket{1}$ be a unit vector such that $\rho=\ketbra{\psi}{\psi}$, and let $\ket{\psi_t}=e^{-itH}\ket{\psi}$. The curve $\rho_t$ is periodic with period $\tau=2\pi/(\epsilon_1-\epsilon_0)$, and according to Eq.\ \eqref{eq: holonomy in terms of lift}, the holonomy of $\rho_t$ is 
\begin{equation}\label{eq: tjufyra}
    \begin{split}
        \hol[\rho_t]
        &=\braket{\psi}{\psi_\tau}\exp\bigg(-\int_0^\tau\braket{\psi_t}{\dot\psi_t}\,\dt\bigg)\\
        &=(|a|^2e^{-i\epsilon_0\tau}+|b|^2e^{-i\epsilon_1\tau})e^{i\tau(\epsilon_0|a|^2+\epsilon_1|b|^2)} \\
        &=e^{2\pi i|b|^2}.
    \end{split}
\end{equation}
Furthermore, the evolution has the speed 
\begin{equation}
    \sqrt{\tfrac{1}{2}\tr(\dot\rho_t^2)}=(\epsilon_1-\epsilon_0)|a||b|
\end{equation}
and, thus, the length 
\begin{equation}
    \length[\rho_t]=\tau(\epsilon_1-\epsilon_0)|a||b|= 2\pi|a||b|.
\end{equation}
Let $\theta$ be the principal argument of the holonomy. Then,
\begin{equation}\label{eq: tjutju}
    \length[\rho_t]^2=2\pi|b|^2(2\pi-2\pi|b|^2)=|\theta|(2\pi-|\theta|).
\end{equation}
We conclude that a qubit with time-independent Hamiltonian saturates the isoholonomic inequality.
\end{ex}

\subsection{The isoholonomic inequality for gates}
In this section, we show that the length of a closed curve in $\GG(n;\HH)$ with holonomy $\Gamma$ is bounded from below by $L(\Gamma)$ as defined in Eq.\ \eqref{eq: the isoholonomic bound}. Then, in Sec.\ \ref{sec: Tightness of the isoholonomic bound}, we show, inspired by Ref.\ \cite{TaNaHa2005}, that $L(\Gamma)$ is a tight bound when the dimension of $\HH$ is greater than or equal to $2n-k$, where $k$ is the number of $1$s in the spectrum of $\Gamma$. We do this by constructing a Hamiltonian that drives $\RR$ in a loop with holonomy $\Gamma$ and length $L(\Gamma)$. The question of whether $L(\Gamma)$ is a tight bound when the dimension of $\HH$ is less than $2n-k$ is still open.

Assume $P_t$ is a closed curve of rank $n$ orthogonal projection operators at $P_\RR$ having holonomy $\Gamma$.
Since ``length'' and ``holonomy'' are parametrization invariant quantities, we can assume that $P_t$ has a constant speed and returns to $P_\RR$ at time $\tau=1$.

Let $e^{i\theta_1},e^{i\theta_2},\dots,e^{i\theta_n}$ be the eigenvalues of $\Gamma$,
with $\theta_j$ being the principal argument of the $j$th eigenvalue.
Let $F$ be a frame consisting of eigenvectors of $\Gamma$,
\begin{equation}\label{eq: frame of eigenvectors}
    F=\big(\ket{u_1}\,\ket{u_2}\dots\ket{u_n}\big),
    \quad
    \Gamma\ket{u_j}=e^{i\theta_j}\ket{u_j},
\end{equation}
and let $F_t$ be the horizontal lift of $P_t$ starting at $F$,
\begin{equation}
    F_t=\big(\ket{u_{1;t}}\,\ket{u_{2;t}}\dots\ket{u_{n;t}}\big),
    \quad
    \ket{u_{j;0}}=\ket{u_j}.
\end{equation}
Since the Stiefel-Grassmann bundle is a Riemannian submersion and $P_t$ has a constant speed, so does $F_t$, and the square of the length of $P_t$ is
\begin{equation}\label{eq: length squared}
    \length[P_t]^2 
    = 2\energy[P_t] 
    = 2\energy[F_t] 
    = 2\sum_{j=1}^n \energy(\ket{u_{j;t}}).
\end{equation}
Furthermore, since $F_t$ is horizontal, each curve $\ket{u_{j;t}}$ is Aharonov-Anandan horizontal,
\begin{equation}\label{eq: component is horizontal}
    \braket{u_{j;t}}{\dot u_{j;t}}=\bra{u_j}FF_t^\dagger \dot F_tF^\dagger\ket{u_j}=0,
\end{equation}
and $\ket{u_{j;t}}$ projects onto a closed curve of pure states $\rho_{j;t}$ with Aharonov-Anandan geometric phase $\theta_j$,
\begin{equation}\label{eq: component has AA-phase}
    \ket{u_{j;1}}=F_1F^\dagger\ket{u_j}=\Gamma\ket{u_j}=e^{i\theta_j}\ket{u_j}.
\end{equation}
The curves $\ket{u_{j;t}}$ and $\rho_{j;t}$ have the same kinetic energies, and by the isoholonomic inequality for states \eqref{eq: the isoholonomic inequality for states}, the length of $\rho_{j;t}$ is lower bounded by $L(\theta_j)$. Thus,
\begin{equation}\label{eq: estimating length}
        2\energy[\ket{u_{j;t}}]
        = 2\energy[\rho_{j;t}]
        \geq L(\theta_j)^2.
\end{equation}
Equations \eqref{eq: length squared} and \eqref{eq: estimating length} imply that 
\begin{equation}
    \length[P_t]^2
    \geq \sum_{j=1}^n L(\theta_j)^2
    = L(\Gamma)^2.
\end{equation}
This proves the isoholonomic inequality \eqref{eq: the isoholonomic inequality}.

\begin{ex}
The quantum Fourier transform 
\begin{equation}
    F_n\ket{u_j}=\frac{1}{\sqrt{n}} \sum_{k=0}^{n-1}e^{2\pi ijk}\ket{u_k}
\end{equation}
is used in many quantum algorithms \cite{NiCh2010}.
The quantum Fourier transform has characteristic polynomial 
\begin{equation}
    (x-1)^{\lfloor\frac{n+4}{4}\rfloor}(x+1)^{\lfloor\frac{n+2}{4}\rfloor}(x+i)^{\lfloor\frac{n+1}{4}\rfloor}(x-i)^{\lfloor\frac{n-1}{4}\rfloor},
\end{equation}
from which we can read off the eigenvalues of the transform and their multiplicities.
We conclude that the Fourier transform has isoholonomic bound
\begin{equation}
    L(F_n)=\pi\sqrt{\Big\lfloor\frac{n+2}{4}\Big\rfloor + \frac{3}{4}\Big(\Big\lfloor\frac{n+1}{4}\Big\rfloor+\Big\lfloor\frac{n-1}{4}\Big\rfloor\Big)}.
\end{equation}
For $n=2$, the Fourier transform equals the Hadamard gate $H$.
The Hadamard gate thus has isoholonomic bound $L(H)=\pi$. We have listed the isoholonomic bounds for a universal set of qubit gates in Table \ref{tab: gates}.
\end{ex}

\subsection{The runtime bound}\label{sec: The Mandelstam-Tamm type bound}
The evolution time estimate $\tau\geq\qsl[H_t;\Gamma]$ follows immediately from the isoholonomic inequality for gates and the observation that if $\RR$ is transported in a loop $\RR_t$ by the Hamiltonian $H_t$, and $P_t$ is the corresponding curve of projection operators, the square of the speed of $\RR_t$ equals the skewness of $H_t$ relative to $\RR_t$,
\begin{equation}
    \gG(\dot P_t,\dot P_t)
    = \frac{1}{2}\tr\big((-i[H_t,P_t])^2\big)=I(H_t; \RR_t).
\end{equation}

If the Hamiltonian is time-independent, $H_t=H$, the skewness is conserved and the speed is constant,
\begin{equation}
\begin{split}
    2I(H; \RR_t)
    &= \tr\big((-i[H,e^{-itH}P_\RR e^{itH} ])^2\big) \\
    &= \tr\big(e^{-itH}(-i[H,P_\RR])^2e^{itH}\big) \\
    &= 2I(H; \RR).
\end{split}
\end{equation}
For a parallel transporting Hamiltonian, we have that
\begin{equation}
    I(H_t;\RR_t)=\tr(H_t^2P_t).
\end{equation}

\section{Time-optimal universal gates}\label{sec: Nonadiabatic holonomic computation}
In the standard description of nonadiabatic holonomic quantum computation \cite{ZhKyFiKwSjTo2023, SjMoCa2016}, input states are prepared in a space $\RR$ associated with a register of qubits. The states are then manipulated with holonomic gates implemented by parallel transporting Hamiltonians.

Typically, a frame of product vectors 
\begin{equation}
    F=\big(\ket{100\cdots 0}\, \ket{010\cdots 0}\dots\ket{111\cdots 1}\big)
\end{equation}
is used as a reference frame in $\RR$ where, at each position, $\ket{0}$ and $\ket{1}$ are orthonormal vectors that span the marginal Hilbert space of the corresponding qubit. The space $\RR$ is referred to as the computational space, $F$ is the computational basis, and states and gates are represented as matrices relative to $F$.

A universal set of quantum gates can approximate any other quantum gate to any desired precision. For a computational system manipulating qubits, the one-qubit Hadamard gate, phase gate, $\pi/8$ gate, and two-qubit CNOT gate form a universal set \cite{NiCh2010}. The isoholonomic bounds for these are listed in Table \ref{tab: gates}.
\begin{table}[t]
    \centering
    \begin{tabular}{l l l}
    \toprule
    Gate & \hspace{3pt}\hspace{3pt}Matrix representation & \hspace{3pt} Isoholonomic bound \\
    \midrule\noalign{\smallskip}
    Hadamard 
    & \hspace{3pt} $H = \frac{1}{\sqrt{2}}\begin{pmatrix} 1 & 1 \\ 1 & -1\end{pmatrix}$ 
    & \hspace{3pt} $L(H)=\pi$ \\
    \noalign{\smallskip}\noalign{\smallskip}
    phase gate
    & \hspace{3pt} $S = \begin{pmatrix} 1 & 0 \\ 0 & i \end{pmatrix}$
    & \hspace{3pt} $L(S)=\frac{\pi\sqrt{3}}{2}$ \\
    \noalign{\smallskip}
    \noalign{\smallskip}
    $\pi/8$ gate
    & \hspace{3pt} $T = \begin{pmatrix} 1 & 0 \\ 0 & e^{\frac{i\pi}{4}} \end{pmatrix}$
    & \hspace{3pt} $L(T)=\frac{\pi\sqrt{7}}{4}$ \\
    \noalign{\smallskip}\noalign{\smallskip}
    CNOT 
    & \hspace{3pt} $\mathit{CNOT} = \begin{pmatrix} 
    1 & 0 & 0 & 0 \\ 
    0 & 1 & 0 & 0 \\
    0 & 0 & 0 & 1 \\
    0 & 0 & 1 & 0 \\
    \end{pmatrix}$ 
    & \hspace{3pt} $L(\mathit{CNOT})=\pi$ \\
    \noalign{\smallskip}
    \bottomrule
    \end{tabular}
    \caption{Isoholonomic bounds for a complete set of qubit gates: the one-qubit Hadamard gate $H$, phase gate $S$, $\pi/8$ gate $T$, and the two-qubit CNOT gate.}
    \label{tab: gates}
\end{table}

Reference \cite{SjToAnHeJoSi2012} by Sjöqvist \emph{et al.}\ contains proposals for how to holonomically implement the one-qubit gates 
\begin{equation}
	\Gamma_1(\alpha,\beta)=
	\begin{pmatrix} \cos\alpha & e^{-i\beta}\sin\alpha  \\ e^{i\beta}\sin\alpha & -\cos\alpha \end{pmatrix}
\end{equation}
and the two-qubit gates
\begin{equation}
    \Gamma_2(\alpha,\beta)=\begin{pmatrix}
        \cos\alpha & 0 & & 0 & e^{-i\beta} \sin\alpha \\
        0 & 1 & & 0 & 0 \\
        0 & 0 & & 1 & 0 \\
        e^{i\beta} \sin\alpha  & 0 & & 0 & -\cos\alpha
    \end{pmatrix},
\end{equation}
which together form a universal set \cite{SjToAnHeJoSi2012, BrDaDoGiHaMoNiOs2002}. We demonstrate below that the proposals in Ref.\ \cite{SjToAnHeJoSi2012} are time optimal in the sense that the length of the trajectory of the computational space equals the isoholonomic bound of the implemented quantum gate.

\subsection{One-qubit gates}
\begin{figure}[t]
    \centering
    \includegraphics[width=.32\textwidth]{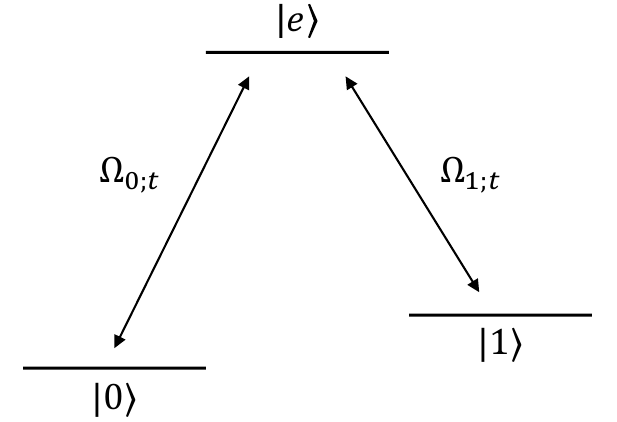}
    \caption{A three-level system in a $\Lambda$ configuration with two closely spaced levels $\ket{0}$ and $\ket{1}$ coupled to an excited level $\ket{e}$ by resonant pulsed laser beams with Rabi frequencies $\Omega_{0;t}$ and $\Omega_{1;t}$, respectively.}
    \label{fig: lambda}
\end{figure}

Following Sjöqvist \emph{et al.}\ \cite{SjToAnHeJoSi2012} we consider a system with three bare energy levels in a $\Lambda$ configuration. 
We assume that the lower, closely spaced levels are represented by $\ket{0}$ and $\ket{1}$
and that the excited energy level, whose energy we set to $0$, is represented by $\ket{e}$; see Fig.\ \ref{fig: lambda}.
Furthermore, we assume that resonant laser pulses drive the transitions $\ket{0}\leftrightarrow\ket{e}$ and 
$\ket{1}\leftrightarrow\ket{e}$ and that the dipole and rotating wave approximations are applicable.
In the rotating frame of the laser fields, the Hamiltonian can then be written
\begin{equation}
    H_t=\Omega_{0;t}\ketbra{e}{0}+\Omega_{0;t}^*\ketbra{0}{e}+\Omega_{1;t}\ketbra{e}{1}+\Omega_{1;t}^*\ketbra{1}{e}.
\end{equation}

We take the sum of the lower energy levels as the computational space and $\ket{0}$ and $\ket{1}$ as the computational basis. 
If the laser pulses have a common envelope, 
\begin{equation}
	\Omega_{j;t}=\Omega(t)\omega_j,\quad |\omega_0|^2+|\omega_1|^2=1,
\end{equation}
the Hamiltonian is parallel transporting, and if the support of the envelope is $[0,\tau]$ and
\begin{equation}
	\int_0^\tau\Omega(t)\,\dt=\pi,
\end{equation}
the Hamiltonian drives the computational space in a loop $\RR_t$ in time $\tau$. The holonomy of the loop is 
\begin{equation}
	\Gamma[\RR_t]=\begin{pmatrix} |\omega_1|^2-|\omega_0|^2 & -2\omega_0^*\omega_1 \\ -2\omega_0\omega_1^* & |\omega_0|^2-|\omega_1|^2 \end{pmatrix},
\end{equation}
and if we adjust the laser pulses so that $\omega_0 = \sin(\alpha/2)e^{i\beta/2}$ and $\omega_1 = -\cos(\alpha/2)e^{-i\beta/2}$,
then $\Gamma[\RR_t]=\Gamma_1(\alpha,\beta)$.

The gate $\Gamma_1(\alpha,\beta)$ has eigenvalues $1$ and $-1$, and hence the isoholonomic bound $L(\Gamma_1(\alpha,\beta))=\pi$.
The loop of the computational space has the length 
\begin{equation}
    \tau\llangle \sqrt{I(H_t;\RR_t)}\,\rrangle = \tau\llangle\Omega(t)\rrangle=\pi.
\end{equation}
Since the length and the isoholonomic bound agree, the implementation is time optimal.

\begin{rmk}
For $\alpha=\pi/4$ and $\beta=0$, the above scheme implements the Hadamard gate time optimally.
However, it cannot generate the phase and $\pi/8$ gates. References \cite{XuLiZhTo2015, Sj2016} contain proposals on how to generate these gates in a $\Lambda$ system with off-resonant driving. Strictly speaking, the Hamiltonians in these proposals are not parallel transporting as they give rise to (irrelevant) dynamical phases. This issue will be addressed in a forthcoming paper.
\end{rmk}

\begin{rmk}
For high-speed computations, the rotating wave approximation may not be valid. Reference \cite{AlSj2022} considers the scheme in Ref.\ \cite{SjToAnHeJoSi2012} without the rotating wave approximation. The proposals in Ref.\ \cite{AlSj2022} are also time optimal. 
\end{rmk}
\vspace{-10pt}

\subsection{Two-qubit gates}
To implement the two-qubit gate $\Gamma_2(\alpha,\beta)$,
Sjöqvist \emph{et al.}\ \cite{SjToAnHeJoSi2012} consider an array of ions, each of which exhibits an internal $\Lambda$ structure as in Fig.\ \ref{fig: lambda}.
The laser pulses that drive the transitions $\ket{j}\leftrightarrow\ket{e}$ are controlled so that the dynamics of a pair of ions is governed by an effective Hamiltonian of the form $H_t=\Omega(t)(H_0+H_1)$, where
\begin{align}
    H_0 &= \omega_{00}\ketbra{ee}{00}+\omega_{11}\ketbra{ee}{11} + \mathrm{H.c.},\\
    H_1 &=\omega_{0e}\ketbra{e0}{0e}+\omega_{1e}\ketbra{e1}{1e} + \mathrm{H.c.}
\end{align}
We take $\ket{00},\ket{01},\ket{10},\ket{11}$ as the computational basis and the span of these vectors as the computational space, and we adjust the
laser pulses so that
\begin{align}
    \omega_{00} &= \sin(\alpha/2)e^{i\beta/2}, \\
    \omega_{11} &= -\cos(\alpha/2)e^{-i\beta/2}, \\
    \omega_{0e} &= \sin(\alpha/2), \\
    \omega_{1e} &= -\cos(\alpha/2),
\end{align}
and so that the envelope function satisfies
\begin{equation}
    \int_0^\tau\Omega(t)\,\dt=\pi.
\end{equation}
The Hamiltonian then parallel transports the computational space in a loop, $\RR_t$, in time $\tau$ and thereby implements the gate $\Gamma_2(\alpha,\beta)$. This gate has eigenvalues $1$ of multiplicity $3$ and $-1$ of multiplicity $1$, and thus the isoholonomic bound $L(\Gamma_2(\alpha,\beta))=\pi$.
Since the length of the loop of the computational space agrees with this bound,
\begin{equation}
    \tau\llangle \sqrt{I(H_t;\RR_t)}\,\rrangle = \tau\llangle\Omega(t)\rrangle=\pi,
\end{equation}
the implementation is time optimal.

\section{Tightness of the isoholonomic bound}\label{sec: Tightness of the isoholonomic bound}
We can generalize the qubit example in Sec.\ \ref{sec: The isoholonomic inequality} to a proof that $L(\Gamma)$ is tight when the dimension of $\HH$ is at least $2n-k$, where $k$ is the number of $1$s in the spectrum of $\Gamma$.
To do this, we arrange the eigenvalues of $\Gamma$ so that
\begin{equation}\label{eq: nollor}
    \theta_{n-k+1} = \theta_{n-k+2} = \dots = \theta_{n} = 0.
\end{equation}
Let $F$ be a frame for $\RR$ of eigenvectors of $\Gamma$ as shown in Eq.\ \eqref{eq: frame of eigenvectors}, and let $\ket{v_1},\ket{v_2},\dots,\ket{v_{n-k}}$ be $n-k$ pairwise orthogonal unit vectors in the orthogonal complement of $\RR$. Within the span of $\ket{u_j}$ and $\ket{v_j}$ choose two orthonormal vectors $\ket{0_j}$ and $\ket{1_j}$, where the latter vector satisfies the condition
\begin{equation}\label{eq: requirement}
    2\pi|\braket{1_j}{u_j}|^2 =\theta_j\mod 2\pi.
\end{equation}
Let $\epsilon_0<\epsilon_1$, and define a Hamiltonian $H$ as
\begin{equation}\label{eq: the Hamiltonian}
    H=\sum_{j=1}^{n-k} \epsilon_0\ketbra{0_j}{0_j} + \epsilon_1\ketbra{1_j}{1_j}.
\end{equation}
Furthermore, define 
\begin{equation}
    F_t = \big(\ket{u_{1;t}}\,\ket{u_{2;t}}\dots\ket{u_{n;t}}\big)
\end{equation}
as $F_t=e^{-itH}F$ and let $P_t=F_tF_t^\dagger$.

For $j=1,2,\dots,n-k$, the vector $\ket{u_j}$ rotates within the span of $\ket{u_j}$ and $\ket{v_j}$ and returns to $\RR$ for the first time at $\tau=2\pi/(\epsilon_1-\epsilon_0)$; for $j=n-k+1,n-k+2,\dots,n$, the vector $\ket{u_j}$ is held fixed. Thus, $\RR$ is driven in a loop $\RR_t$ with period $\tau=2\pi/(\epsilon_1-\epsilon_0)$.

According to Eq.\ \eqref{eq: holonomy in terms of lift}, the holonomy of $\RR_t$ is 
represented by the matrix
\begin{equation}
    F^\dagger \hol[\RR_t] F 
    = F^\dagger F_\tau \overset{_\leftarrow}{\mathcal{T}} \exp \bigg( -\int_0^\tau F_t^\dagger\dot F_t\,\dt \bigg)
\end{equation}
relative to $F$. Since the 
$\ket{u_j}$s rotate in pairwise perpendicular subspaces, $F^\dagger F_\tau$ and $F_t^\dagger \dot F_t$ are diagonal matrices,
\begin{align}
    F^\dagger F_\tau &= \diag\big(\braket{u_1}{u_{1;\tau}}, \dots, \braket{u_n}{u_{n;\tau}}\big),\\
    F_t^\dagger \dot F_t &= \diag\big(\braket{u_{1;t}}{\dot u_{1;t}}, \dots, \braket{u_{n;t}}{\dot u_{n;t}}\big).
\end{align}
For $j=1,2,\dots,n-k$ write $\ket{u_j}=a_j\ket{0_j}+b_j\ket{1_j}$. Then, as in Eq.\ \eqref{eq: tjufyra},
\begin{align}
    \braket{u_j}{u_{j;\tau}}&=|a_j|^2e^{-i\epsilon_0\tau}+|b_j|^2e^{-i\epsilon_1\tau},\\
    \braket{u_{j;t}}{\dot u_{j;t}}&=-i(\epsilon_0|a_j|^2+\epsilon_1|b_j|^2).
\end{align}
Also, $\braket{u_j}{u_{j;\tau}}=1$ and $\braket{u_{j;t}}{\dot u_{j;t}}=0$
for $j=n-k+1,n-k+2,\dots,n$. We conclude that 
\begin{equation}
    F^\dagger \hol[\RR_t] F 
    =\diag(e^{2\pi i|b_1|^2},\dots,e^{2\pi i|b_{n-k}|^2},1,\dots,1)
\end{equation}
which, by assumptions \eqref{eq: nollor} and \eqref{eq: requirement}, shows that the holonomy of $\RR_t$ is $\Gamma$.

To calculate the length of $\RR_t$ we write $\rho_{j;t}=\ketbra{u_{j;t}}{u_{j;t}}$ and observe that
the square of the speed of $\RR_t$ is 
\begin{equation}
        \frac{1}{2}\tr\big(\dot P_t^2 \big)
        = \sum_{j=1}^{n} \frac{1}{2}\tr\big(\dot \rho_{j;t}^2 \big)
        = \sum_{j=1}^{n-k} (\epsilon_1-\epsilon_0)^2|a_j|^2|b_j|^2.
\end{equation}
Since the speed is constant, the length of $\RR_t$ squared is 
\begin{equation}
    \begin{split}
        \tau^2\sum_{j=1}^{n-k} \frac{1}{2}\tr\big(\dot \rho_{j;t}^2 \big)
        = \sum_{j=1}^{n-k} 4\pi^2|a_j|^2|b_j|^2
        = \sum_{j=1}^{n-k} L(\theta_j)^2.
    \end{split}
\end{equation}
The second identity follows from Eq.\ \eqref{eq: tjutju} and the assumption \eqref{eq: requirement}.
We conclude that $\RR_t$ has length $L(\Gamma)$.

\begin{rmk}
The calculations above show that if the dimension of $\RR$ is at most half the dimension of $\HH$, and every direct sum of qubit Hamiltonians can be generated, then every gate on $\RR$ can be implemented time optimally.    
\end{rmk}

\begin{rmk}
The Hamiltonian in Eq.\ \eqref{eq: the Hamiltonian} need not be parallel transporting. 
To define a parallel transporting Hamiltonian $\bar{H}_t$ that drives $\RR$ along the same 
trajectory as $H$ we use Eq.\ \eqref{eq: parallel for constant H} and define $\bar{H}_t$ as
\begin{equation}
    \bar{H}_t
    = e^{-itH}(HP_\RR + P_\RR H - 2 P_\RR H P_\RR)e^{itH}.
\end{equation}
To simplify the expression for $\bar{H}_t$, we assume the phases of $\ket{0_j}$ and $\ket{1_j}$ are such that $a_j$ and $b_j$ are real. Then,
\begin{equation}
\begin{split}
    &\bar{H}_t
    = (\epsilon_1-\epsilon_0)\sum_{j=1}^{n-k} a_jb_j\Big(2a_jb_j\big(\ketbra{1_j}{1_j}-\ketbra{0_j}{0_j}\big) \\
    &+ (a_j^2-b_j^2)\big(e^{it(\epsilon_1-\epsilon_0)}\ketbra{0_j}{1_j}+e^{it(\epsilon_0-\epsilon_1)}\ketbra{1_j}{0_j}\big)\Big). 
\end{split}
\end{equation}
\end{rmk}

\section{Summary}
We have derived an estimate, called the isoholonomic inequality, for the length of a cyclic transformation of a subspace of a Hilbert space with a given holonomy. The isoholonomic inequality constitutes half of the solution to the isoholonomic problem for holonomic quantum gates, as formulated in Ref.\ \cite{Mo1990} (see Ref.\ \cite{TaNaHa2005} for the other half). We have also converted the isoholonomic inequality into an estimate of the time required to execute a holonomic quantum gate unitarily. As an illustration, we have shown that the implementation scheme in Ref.\ \cite{SjToAnHeJoSi2012} for a universal set of holonomic gates is time-optimal. The paper ended with a proof that the isoholonomic inequality is tight if the dimension of the subspace being transformed is at most half of the dimension of the Hilbert space.

\section*{Acknowledgments}
The author thanks Niklas Hörnedal for fruitful discussions and for proofreading early drafts.

\appendix

\titleformat{\section}[block]{\bfseries\large}{Appendix \Alph{section}:}{0.9em}{}
\titlespacing{\section}{0em}{1.2em}{0.7em}

\section{Derivation of the isoholonomic inequality for states}\label{app: Derivation of the isoholonomic inequality for states}
Let $L(\theta)$ be the shortest length a closed curve of pure states can have, given that its Aharonov-Anandan holonomy is $e^{i\theta}$, where $-\pi<\theta\leq \pi$. We show that 
\begin{equation}
    L(\theta)=\sqrt{|\theta|(2\pi-|\theta|)}.
\end{equation} 
Since there is nothing to prove if $\theta=0$, we assume $\theta\ne 0$. 

Let $\rho$ be an arbitrary pure state, and let $\rho_t$ be a closed curve of pure states at $\rho$ with holonomy $e^{i\theta}$ and length $L(\theta)$.\footnote{$L(\theta)$ does not depend on the choice of initial state $\rho$ because length and Aharonov-Anandan holonomy are unitarily invariant quantities. Furthermore, each $\theta$ in $(-\pi,\pi]$ is the Aharonov-Anandan geometric phase of a closed curve of pure states at $\rho$, and at least one of these has length $L(\theta)$ since $\GG(1;\HH)$ is compact.}
Since a reparameterization of $\rho_t$ does not change its length and holonomy, we can assume that $\rho_t$ has a constant speed and returns to its initial state $\rho$ at $\tau=1$.

Let $\ket{\psi}$ be a unit vector projecting onto $\rho$, and let $\ket{\psi_t}$ be the horizontal lift of $\rho_t$ starting from $\ket{\psi}$. The curve $\ket{\psi_t}$ extends from $\ket{\psi}$ to $e^{i\theta}\ket{\psi}$, both of which project onto $\rho$. Since the Hopf bundle is a Riemannian submersion, $\ket{\psi_t}$ and $\rho_t$ have the same length. Furthermore, $\ket{\psi_t}$ has the same constant speed as $\rho_t$. Thus, 
\begin{equation}
    L(\theta)=\int_0^1\sqrt{\braket{\dot\psi_t}{\dot\psi_t}}\,\dt=\sqrt{\braket{\dot\psi_0}{\dot\psi_0}}.
\end{equation}

The curve $\ket{\psi_t}$ is an extremal for the augmented kinetic energy functional
\begin{equation}
    \energy\big[\ket{\phi_t},\lambda_t\big]
    =\frac{1}{2} \int_0^1 \big( \braket{\dot\phi_t}{\dot\phi_t}+2i\lambda_t\braket{\phi_t}{\dot\phi_t}\big) \,\dt,
\end{equation}
with $\lambda_t$ being a Lagrange multiplier that forces horizontality. This is because $\rho_t$ is a closed curve of minimal length among those having holonomy $e^{i\theta}$ \cite{Mo1990}. The kinetic energy functional is defined on the space of curves in $\VV(1;\HH)$ extending from $\ket{\psi}$ to $e^{i\theta}\ket{\psi}$ over the time interval $[0,1]$.

Each variational vector field of $\ket{\psi_t}$ that fixes the endpoints of $\ket{\psi_t}$ has the form $-iX_t\ket{\psi_t}$, where $X_t$ is an arbitrary curve of Hermitian operators that vanishes for $t=0$ and $t=1$. A variation of $\ket{\psi_t}$ with this variation vector field is $\ket{\psi_{\epsilon,t}}=U_{\epsilon,t}\ket{\psi_t}$ where, for each $\epsilon$ in $(-1,1)$, $U_{\epsilon,t}$ is the backward time-ordered exponential of $-i\epsilon \dot X_t$,
\begin{equation}
    U_{\epsilon,t}=\overset{_\rightarrow}{\mathcal{T}}\exp\bigg(-i\epsilon\int_0^t \dot X_{s}\,\d s\bigg).
\end{equation}
A partial integration shows that the variational derivative of the augmented kinetic energy of the variation of $\ket{\psi_t}$ is
\begin{widetext}
\begin{ceqn}
\begin{equation}\label{partialintegration}
        \frac{\d}{\d \epsilon}\energy\big[\ket{\psi_{\epsilon,t}},\lambda_t\big]\Big|_{\epsilon=0}
   = \frac{1}{2} \int_0^1 \tr \bigg( X_t\frac{\d}{\d t}\Big(i\ketbra{\psi_t}{\dot\psi_t} -i\ketbra{\dot\psi_t}{\psi_t} -2\lambda_t\ketbra{\psi_t}{\psi_t}\Big)\bigg)\dt.    
\end{equation} 
\end{ceqn}
\end{widetext}
The Lagrange equation for $\ket{\psi_t}$ thus reads 
\begin{equation}\label{eq: Lagrange equation}
 \frac{\d}{\d t}\big(i\ketbra{\psi_t}{\dot\psi_t}-i\ketbra{\dot\psi_t}{\psi_t}-2\lambda_t\ketbra{\psi_t}{\psi_t}\big)=0.
\end{equation}
From the Lagrange equation follows that
\begin{equation}\label{eq: A}
    A=i\ketbra{\psi_t}{\dot\psi_t}-i\ketbra{\dot\psi_t}{\psi_t}-2\lambda_t\ketbra{\psi_t}{\psi_t} 
\end{equation}
is a time-independent Hermitian operator, and from Eq.\ \eqref{eq: A} follows that the Lagrange multiplier is time-independent, $    2\dot\lambda_t=\bra{\dot\psi_t}A\ket{\psi_t}+\bra{\psi_t}A\ket{\dot\psi_t}=0$.
We write $\lambda$ for the value of the Lagrange multiplier. By Eq.\ \eqref{eq: A},
\begin{equation}\label{eq: psi dynamics}
    \ket{\dot\psi_t}=-i(A-2\lambda)\ket{\psi_t}.
\end{equation}
Equation \eqref{eq: A} also tells us that the support of $A$ is two-dimensional and is spanned by $\ket{\psi}$ and $\ket{\dot\psi_0}$. 
From this observation and Eq.\ \eqref{eq: psi dynamics} we can conclude that $\ket{\psi}$, and thus the entire curve $\ket{\psi_t}$, is contained in the sum of two eigenspaces of $A-2\lambda$. The eigenvalues are
\begin{equation}
    a_{\pm}=-\lambda\pm\sqrt{\lambda^2+\braket{\dot\psi_0}{\dot\psi_0}}.
\end{equation}
The holonomy condition $\ket{\psi_1}=e^{i\theta}\ket{\psi}$ is satisfied if and only if 
$a_+=2\pi k-\theta$ for an integer $k\geq0$ and $a_-=2\pi l-\theta$ for an integer $l\leq 0$. Hence,
\begin{equation}\label{1D length estimate}
\begin{split}
    L(\theta)^2
    &=\braket{\dot\psi_0}{\dot\psi_0} \\
    &=-a_+a_- \\
    &=-(2\pi k-\theta)(2\pi l-\theta) \\
    &\geq |\theta|(2\pi-|\theta|).
\end{split}  
\end{equation}
That the last inequality is an equality follows, for example, from the observation that every cyclic evolution of a qubit generated by a time-independent Hamiltonian saturates the inequality, as was shown in Sec.\ \ref{sec: The isoholonomic inequality}.

\twocolumngrid


\begin{thebibliography}{99}
\bibitem{WiZe1984}
F. Wilczek and A. Zee, 
Appearance of gauge structure in simple dynamical systems,
\href{https://doi.org/10.1103/PhysRevLett.52.2111}{Phys. Rev. Lett. {52}, 2111 (1984)}.

\bibitem{ZaRa1999}
P. Zanardi and M. Rasetti,
Holonomic quantum computation, 
\href{https://doi.org/10.1016/S0375-9601(99)00803-8}{Phys. Lett. A {264}, 94 (1999)}.

\bibitem{PaZaRa1999}
J. Pachos, P. Zanardi, and M. Rasetti,
Non-Abelian Berry connections for quantum computation,
\href{https://doi.org/10.1103/PhysRevA.61.010305}{Phys. Rev. A {61}, 010305(R) (1999)}.

\bibitem{PaZa2001}
J. Pachos and P. Zanardi,
Quantum holonomies for quantum computing,
\href{https://doi.org/10.1142/S0217979201004836}{Int. J. Mod. Phys. B {15}, 1257 (2001)}.

\bibitem{An1988}
J. Anandan,
Non-adiabatic non-abelian geometric phase,
\href{https://doi.org/10.1016/0375-9601(88)91010-9}{Phys. Lett. A, {133}, 171 (1988)}.

\bibitem{SjToAnHeJoSi2012}
E. Sj\"oqvist, D. M. Tong, L. M. Andersson, B. Hessmo, M. Johansson, and K. Singh,
Non-adiabatic holonomic quantum computation,
\href{https://doi.org/10.1088/1367-2630/14/10/103035}{New J. Phys. {14}, 103035 (2012)}.

\bibitem{XuZhToSjKw2012}
G. F. Xu, J. Zhang, D. M. Tong, E. Sj\"oqvist, and L. C. Kwek,
Nonadiabatic holonomic quantum computation in decoherence-free subspaces,
\href{https://doi.org/10.1103/PhysRevLett.109.170501}{Phys. Rev. Lett. {109}, 170501 (2012)}.

\bibitem{AlSj2022}
G. O. Alves and E. Sj\"oqvist,
Time-optimal holonomic quantum computation,
\href{https://doi.org/10.1103/PhysRevA.106.032406}{Phys. Rev. A {106}, 032406 (2022)}.

\bibitem{ZhKyFiKwSjTo2023}
J. Zhang, T. H. Kyaw, S. Filipp, L.-C. Kwek, E. Sjöqvist, and D. Tong,
Geometric and holonomic quantum computation,
\href{https://doi.org/10.1016/j.physrep.2023.07.004}{Phys. Rep. {1027}, 1 (2023)}.

\bibitem{Wang2001}
W. Xiang-Bin and M. Keiji,
Nonadiabatic conditional geometric phase shift with NMR,
\href{https://doi.org/10.1103/PhysRevLett.87.097901}{Phys. Rev. Lett. {87}, 097901 (2001)}.

\bibitem{Zhu2002}
S.-L. Zhu and Z. D. Wang,
Implementation of universal quantum gates based on nonadiabatic geometric phases,
\href{https://doi.org/10.1103/PhysRevLett.89.097902}{Phys. Rev. Lett. {89}, 097902 (2002)}.

\bibitem{HoSo2023c}
N. H\"ornedal and O. S\"onnerborn,
Tight lower bounds on the time it takes to generate a geometric phase, 
\href{https://doi.org/10.1088/1402-4896/acf8a2}{Phys. Scr. {98}, 105108 (2023)}.

\bibitem{TaNaHa2005}
S. Tanimura, M. Nakahara, and D. Hayashi,
Exact solutions of the isoholonomic problem and the optimal control problem in holonomic quantum computation,
\href{https://doi.org/10.1063/1.1835545}{J. Math. Phys. {46}, 022101 (2005)}.

\bibitem{Mo1990}
R. Montgomery,
Isoholonomic problems and some applications,
\href{https://doi.org/10.1007/BF02096874}{Commun. Math. Phys. {128}, 565 (1990)}.

\bibitem{AbFiJuPeBeWaFi2013}
A. A. Abdumalikov, Jr., J. M. Fink, K. Juliusson, M. Pechal, S. Berger, A. Wallraff, and S. Filipp,
Experimental realization of non-Abelian non-adiabatic geometric gates,
\href{https://doi.org/10.1038/nature12010}{Nature (London) {496}, 482 (2013)}.

\bibitem{FeXuLo2013}
G. Feng, G. Xu, and G. Long,
Experimental realization of nonadiabatic holonomic quantum computation,
\href{https://doi.org/10.1103/PhysRevLett.110.190501}{Phys. Rev. Lett. {110}, 190501 (2013)}.

\bibitem{A-CLaHeBa2014}
S. Arroyo-Camejo, A. Lazariev, S. W. Hell, and G. Balasubramanian, 
Room temperature high-fidelity holonomic single-qubit gate on a solid-state spin,
\href{https://doi.org/10.1038/ncomms5870}{Nat Commun. {5}, 4870 (2014)}.

\bibitem{ZuWaHeZhDaWaDu2014}
C. Zu, W.-B. Wang, L. He, W.-G. Zhang, C.-Y. Dai, F. Wang, and L.-M. Duan,
Experimental realization of universal geometric quantum gates with solid-state spins,
\href{https://doi.org/10.1038/nature13729}{Nature (London) {514}, 72 (2014)}.

\bibitem{Gi2014}
D. Girolami,
Observable measure of quantum coherence in finite dimensional systems,
\href{https://doi.org/10.1103/PhysRevLett.113.170401}{Phys. Rev. Lett. {113}, 170401 (2014)}.

\bibitem{LuSu2020}
S. Luo and Y. Sun,
Skew information revisited: Its variants and a comparison of them,
\href{https://doi.org/10.1134/S0040577920010092}{Theor. Math. Phys. {202}, 104 (2020)}. 

\bibitem{KoNo1996}
S. Kobayashi and K. Nomizu,
\emph{Foundations of Differential Geometry}, Vols. I, II, Wiley Classics Library (Wiley, New York, 1996).

\bibitem{AhAn1987}
Y. Aharonov and J. Anandan,
Phase change during a cyclic quantum evolution,
\href{https://doi.org/10.1103/PhysRevLett.58.1593}{Phys. Rev. Lett. {58}, 1593 (1987)}.

\bibitem{YoTo2023}
X.-D. Yu and D. M. Tong,
Evolution operator can always be separated into the product of holonomy and dynamic operators,
\href{https://link.aps.org/doi/10.1103/PhysRevLett.131.200202}{Phys. Rev. Lett. {131}, 200202 (2023)}.

\bibitem{NiCh2010}
M. A. Nielsen and I. L. Chuang,
\emph{Quantum Computation and Quantum Information},
10th Anniversary Edition (Cambridge University Press, New York, 2010).

\bibitem{SjMoCa2016}
E. Sjöqvist, V. Azimi Mousolou, and C. M. Canali,
Conceptual aspects of geometric quantum computation,
\href{https://doi.org/10.1007/s11128-016-1381-1}{Quantum Inf. Process. {15}, 3995 (2016)}.  

\bibitem{BrDaDoGiHaMoNiOs2002}
M. J. Bremner, C. M. Dawson, J. L. Dodd, A. Gilchrist, A. W. Harrow, D. Mortimer, M. A. Nielsen, and T. J.  Osborne,
Practical scheme for quantum computation with any two-qubit entangling gate,
\href{https://doi.org/10.1103/PhysRevLett.89.247902}{Phys. Rev. Lett. {89}, 247902 (2002)}.

\bibitem{Sj2016}
E. Sjöqvist,
Nonadiabatic holonomic single-qubit gates in off-resonant $\Lambda$ systems,
\href{https://doi.org/10.1016/j.physleta.2015.10.006}{Phys. Lett. A {380}, 65 (2016)}.

\bibitem{XuLiZhTo2015}
G. F. Xu, C. L. Liu, P. Z. Zhao, and D. M. Tong, 
Nonadiabatic holonomic gates realized by a single-shot implementation,
\href{https://doi.org/10.1103/PhysRevA.92.052302}{Phys. Rev. A {92}, 052302 (2015)}.

\end{thebibliography}
\end{document}